\documentclass[twoside,12pt]{article}
\newcommand{\beq}{\begin{equation}}
\newcommand{\eeq}{\end{equation}}
\usepackage{epsfig}
\begin{document}

\title{\mbox{ } \\[-3cm]
{\footnotesize\hspace*{\fill}February -- 1998\\
[-4mm] \footnotesize\hspace*{\fill}CFNUL 02/1998} \\[2cm] 
{\bf Axially Symmetric Cosmological Models with Perfect Fluid
and Cosmological Constant}\thanks{
Based on a contributed paper delivered at the
``THE NON-SLEEPING UNIVERSE: FROM GALAXIES TO THE HORIZON"
Conference held at Porto, November 27-29, 1997.}}

\author{{\bf Paulo Aguiar} \thanks{
E-mail: {\tt paguiar@cosmo.cii.fc.ul.pt}} \hspace{.1mm}
{and {\bf Paulo Crawford}} 
\thanks{E-mail: {\tt crawford@cosmo.cii.fc.ul.pt}} \\ [.4cm]
{\em Centro de F\'{\i}sica Nuclear e Dep. F\'{\i}sica da F. C.
da U. de Lisboa} \\
[.2cm] {\em Av. Prof. Gama Pinto, 2 1699 Lisboa Codex, PORTUGAL}}
\date{}
\maketitle

\begin{abstract}

Following recent considerations of a non-zero value for the vacuum
energy density and the realization that a simple Kantowski-Sachs model might fit
the classical tests of cosmology, we study the qualitative behavior of three
anisotropic and homogeneous models: Kantowski-Sachs, Bianchi I and
Bianchi III universes, with dust and cosmological constant, in order to
find out which are physically permitted.

In fact, these models undergo isotropisation, except for
the Kantowski-Sachs model $(\Omega_{k_{0}}>0)$ with $\Omega_{\Lambda_{0}}<
\Omega_{\Lambda_{M}}$ and for the Bianchi III $(\Omega_{k_{0}}<0)$ with
$\Omega_{\Lambda_{0}}<\Omega_{\Lambda_{M}}$, and the observations will not
be able to distinguish between these models and the standard model.

If we impose that the Universe should be very much isotropic since the 
last scattering epoch ($z\approx 1000$), meaning that the Universe should have
approximately the same Hubble parameter in all directions, we are
led to a matter density parameter very close to the unity at the present time.
\vspace*{0.5cm}

{\setlength{\parindent}{0pt} PACS number(s): 98.80.$-$k, 98.80.Es,
98.80.Hw, 04.20.$-$q}
\end{abstract}

\newpage

Lately, the issue of whether or not there is a non-zero value for the
vacuum energy density, or cosmological constant, has been raised quite
often. Even taking the Hubble constant to be in the range 60-75 km/s/Mpc
it is possible to show \cite{Roos} that the standard model of flat space with
vanishing cosmological constant is ruled out. On the other hand, 
if the classical tests of cosmology are applied to a simple
Kantowski-Sachs metric and the results compared with those obtained for the
standard model, the observations will not be able to distinguish
between these models if the Hubble parameters along the
orthogonal directions are assumed to be approximately equal
\cite{Henriques}.
Indeed, as Collins and Hawking \cite{Collins} point out, the number
of cosmological solutions which demonstrate exact isotropy well
after the big bang origin of the Universe is a small fraction of
the set of allowable solutions of the cosmological equations.
It is therefore prudent to take seriously the possibility that
the Universe is expanding anisotropically. Note also that in
\cite{Crawford} it is shown that some shear free anisotropic
models display a FLRW-like behaviour.
Taking all this into consideration, we discuss the behavior of
some homogeneous but anisotropic models with axial symmetry,
filled with a pressureless
perfect fluid (dust) and a non vanishing cosmological constant.
For this, we will restrict our study to the the metric forms
\begin{equation}
ds^2=-c^2dt^2+a^2(t)dr^2+b^2(t)\left(\frac{dv^2}{1-kv^2}
+v^2d\phi^2\right),
\label{metric}
\end{equation}
with the two scale factors $a(t)$ and $b(t)$; $k$ is the curvature
index of the 2-dimensional surface $dv^2/(1-kv^2)+v^2d\phi^2$ and 
can take the values $+1, 0, -1$, giving the following three
different metrics: Kantowski-Sachs, Bianchi I, and Bianchi III,
respectively. 

Einstein equations for the metric (\ref{metric}), for which
the matter content is in the form of a perfect fluid and 
a cosmological term, $\Lambda$, are then as follows:
\begin{equation}
2~\frac{\dot a}{a}~\frac{\dot b}{b}+\frac{\dot b^2}{b^2}+\frac{kc^2}{b^2}
=8\pi G\rho+\Lambda c^2,
\label{EE1a}
\end{equation}
\begin{equation}
2~\frac{\ddot b}{b}+\frac{\dot b^2}{b^2}+\frac{kc^2}{b^2}
=-8\pi G \frac{p^2}{c^2}+\Lambda c^2,
\label{EE1b}
\end{equation}
\begin{equation}
\frac{\ddot a}{a}+\frac{\ddot b}{b}+\frac{\dot a}{a}~\frac{\dot b}{b}
=-8\pi G \frac{p^2}{c^2}+\Lambda c^2,
\label{EE1c}
\end{equation}
where $\rho$ is the matter density and $p$ is the (isotropic) pressure
of the fluid. Here $G$ is the Newton's gravitational constant and $c$
is the speed of light. If we consider a vanishing pressure $(p=0)$,
which is compatible with the present conditions for the Universe,
the last two equations take the form
\begin{equation}
2~\frac{\ddot b}{b}+\frac{\dot b^2}{b^2}+\frac{kc^2}{b^2}
=\Lambda c^2,
\label{EE2b}
\end{equation}
\begin{equation}
\frac{\ddot a}{a}+\frac{\ddot b}{b}+\frac{\dot a}{a}~\frac{\dot b}{b}
=\Lambda c^2,
\label{EE2c}
\end{equation}
and Eq.(5) can easily be integrated to give
\begin{equation}
\frac{\dot b^2}{b^2}=\frac{M_1}{b^3}-\frac{kc^2}{b^2}+
\frac{\Lambda}{3}c^2,
\label{eqC1}
\end{equation}
where $M_1$ is a constant of integration.

The Hubble constants corresponding to the scales $a(t)$ and $b(t)$
are defined by 

$$H_a\equiv\dot a/a \quad \mbox{and} \quad H_b \equiv  \dot b/b.$$

Using them to introduce the following dimensionless parameters,
in analogy wiht which it is usually done in the
Friedmann-Lema\^{\i}tre-Robertson-Walker (FLRW) universes,
let us define
\begin{equation}
\frac{M_1}{b^3 H_b^2}\equiv \Omega_M,
\end{equation}
\begin{equation}
\frac{kc^2}{b^2 H_b^2}\equiv \Omega_k
\end{equation}
and
\begin{equation}
\frac{\Lambda c^2}{3 H_b^2}\equiv \Omega_{\Lambda}.
\end{equation}
The conservation equation (\ref{eqC1}) can now be rewritten
as
\begin{equation}
\Omega_M-\Omega_k+\Omega_{\Lambda}=1.
\label{eqFG2}
\end{equation}

Now defining the dimensionless variable $y=b/b_0$ where $b_0=b(t_0)$,
is the angular scale factor for the present age of the Universe, and
using Eq.(\ref{eqFG2}) (taken for $t=t_0$), one may rewrite
Eq.(\ref{eqC1}) as
\begin{equation}
\dot y=\pm H_{b_{0}}\sqrt{1+\Omega_{M_{0}}(\frac{1}{y}-1)+
\Omega_{\Lambda_{0}}(y^2-1)},
\label{eqy}
\end{equation}
where the density parameters $\Omega$ and $H_b$ with the zero subscript,
denote as before these quantities at the present time $t_0$. In this
way, the number of independent parameters have been reduced.
Substituting Eq.(7) into Eq.(2) gives
\begin{equation}
\dot a=\frac{M_{\rho}-M_1~\frac{a}{b}+\frac{2}{3}\Lambda c^2ab^2}
{2\sqrt{M_1b-kc^2b^2+\frac{\Lambda}{3}c^2b^4}},
\label{dota}
\end{equation}
where $M_{\rho}$ is a constant proportional to the matter in the
Universe,
\begin{equation}
M_{\rho}=8\pi G\rho ab^2. 
\end{equation}
Using the procedure above, Eq.(13) can be rewritten in the form
\begin{equation}
\Omega_{\rho}-\Omega_M+2~\Omega_{\Lambda}=2~\frac{H_a}{H_b},
\label{eqC3}
\end{equation}
where
\begin{equation}
\Omega_{\rho}=\frac{M_{\rho}}{ab^2H_b^2}.
\label{eqC4}
\end{equation}
>From Eq.(2) one may define a matter density parameter
\begin{equation}
\Omega ={{8\pi G\rho } \over {2H_aH_b+H_b^2}},
\end{equation}
just like in FLRW models, and such that $\Omega=1$ when $k=0$ and
$\Lambda=0$, and which is related to $\Omega_{\rho}$ by
\begin{equation}
\Omega ={{\Omega _\rho} \over {1+2{{H_a} \over {H_b}}}}.
\end{equation}
Although $\Omega_M$ is not the matter density parameter, it performs the
same important role.
We emphasize the fact that if for one particular time $H_a=H_b$
and $\Omega_{\Lambda}=1$, then, by Eqs.(11), (15) and (18),
$3\Omega=\Omega_{\rho}=\Omega_{M}=\Omega_{k}$;
and if $ 0<\Omega_{\Lambda}\ll 1$ and $\Omega_M =1$, then,
$\Omega_k=\Omega_{\Lambda}$ and $\Omega\simeq 1$.

Introducing another dimensionless variable $x=a/a_0$, 
Eq.(13) takes the form
\begin{equation}
\dot x=H_{b_{0}}~\frac{\frac{\Omega_{M_{0}}}{2}(1-\frac{x}{y})
+\Omega_{\Lambda_{0}}(-1+xy^2)+\frac{H_{a_{0}}}{H_{b_{0}}}}
{y\sqrt{\Omega_{M_{0}}(\frac{1}{y}-1)+\Omega_{\Lambda_{0}}
(y^2-1)+1}},
\label{eqx}
\end{equation}
and its number of independent parameters was also reduced, now at
the expense of Eq.(15) taken for the present time $t=t_0$.

The behavior of $y(t)$ may be carried out looking for the $y$ values
where $\dot y=0$.
This analysis was made by Mariano Moles \cite{Moles} for FLRW
models, in great detail.

There are two $\Omega_{\Lambda}$ values which characterize
two zones of distinct behavior of scale factor $b$.
Starting with condition $\dot y=0$ one may obtain
\begin{equation}
\Omega_{\Lambda_{0}}=\frac{\Omega_{M_{0}}(y-1)-y}{y^3-y}
\end{equation}
If we consider $\Omega_{\Lambda_{0}}=\Omega_{\Lambda_{0}}(y)$, as
a function of $y$, then 
this function presents a relative maximum and a minimum, that we will
denote by $\Omega_{\Lambda_{c}}$ and $\Omega_{\Lambda_{M}}$, respectively.
The relative maximum depends on $\Omega_{M_{0}}$ in the following way:
For $\Omega_{M_{0}}<1/2$
we have
\begin{eqnarray}
\Omega_{\Lambda_{c}} & = & \frac{3\Omega_{M_{0}}}{2}\left\{ \left[
\sqrt{ \frac{(\Omega_{M_{0}}-1)^2}{\Omega_{M_{0}}^2}-1}+
\frac{1-\Omega_{M_{0}}}{\Omega_{M_{0}}} \right]^{1/3}+
\right. \nonumber
\\ & &
\left.
\frac{1}{\left[ \sqrt{(\Omega_{M_{0}}-1)^2/\Omega_{M_{0}}^2-1
}+(1-\Omega_{M_{0}})/\Omega_{M_{0}} \right]^{1/3}} \right\}-
(\Omega_{M_{0}}-1),
\end{eqnarray}
for $\Omega_{M_{0}}\geq 1/2$ the expression is
\begin{equation}
\Omega_{\Lambda_{c}}=-3\Omega_{M_{0}}\cos\left(
\frac{\theta+2\pi}{3}\right) -(\Omega_{M_{0}}-1).
\end{equation}
The relative minimum is done by
\begin{equation}
\Omega_{\Lambda_{M}}=-3\Omega_{M_{0}}\cos \left(
\frac{\theta+4\pi}{3}\right) -(\Omega_{M_{0}}-1),
\end{equation}
where $\theta=\cos^{-1}\left( \frac{\Omega_{M_{0}}-1}
{\Omega_{M_{0}}} \right)$.
These expressions are limiting zones of the ($\Omega_{\Lambda_{0}},
\Omega_{M_{0}}$) plane, where $\dot y=0$ has three or one solutions
(for details see \cite[Moles]{Moles}).
The $\Omega_{\Lambda_{M}}$ expression is also defined for 
$\Omega_{M_{0}}>1/2$, but it has the meaning of a maximum only for 
$\Omega_{M_{0}}>1$.
The $\Omega_{\Lambda_{0}}$ less or equal to $\Omega_{\Lambda_{M}}$
imposes the recollapse of scale factor
$b$, while greater values produces inflexional behaviors for $b$.
The $\Omega_{\Lambda_{0}}$ values greater or equal
to $\Omega_{\Lambda_{c}}$ are physicaly ``forbidden" because they
don't reproduce the present Universe (see \cite{Moles}).
Obviously, $\Omega_{\Lambda_{M}} < \Omega_{\Lambda_{c}}$ always.

Although we are considering anisotropic models, the Eq.(12) is
exactly the same as the one obtained by \cite{Moles} for the homogeneous
and isotropic FLRW models. From Eq.(12) and Eq.(19) one obtains
the differential equation
\begin{equation}
\frac{dx}{dy}=\frac{\frac{\Omega_{M_{0}}}{2}(1-\frac{x}{y})
+\Omega_{\Lambda_{0}}(-1+xy^2)+\frac{H_{a_{0}}}{H_{b_{0}}}}
{\Omega_{M_{0}}(1-y)+\Omega_{\Lambda_{0}}(y^3-y)+y}.
\end{equation}

This equation has to comply with the conditions imposed by Eq.(11) and
Eq.(15) evaluated at $t_0$. There are some particular values of the parameter
for which this equation has exact solutions. However, for the
majority of the values of the parameter, the solution has only been
obtained by numerical integration.

Although we are dealing with anisotropic models, we may admit that
at a certain moment of time, which we can take as the present time
$t_0$, the Hubble parameters along the orthogonal directions may be
assumed to be approximately equal, $H_a\simeq H_b$. 
This hypothesis has been considered in \cite[Henriques]{Henriques}
for the case of a Kantowski-Sachs (KS) model. From this study was
derived the conclusion that the classical tests of cosmology are
not at present sufficiently accurate to distinguish between a
FLRW model and the KS defined in that paper, with
$(H_{a_{0}}\simeq H_{b_{0}})$, except for small values of $b_0$.

The Eq.(24), with $H_{a_{0}}=H_{b_{0}}$, has three distinct possible
integrations, one for each $k$ value, (see Figure 1).
\begin{figure}[ht]
\centering{\psfig{figure=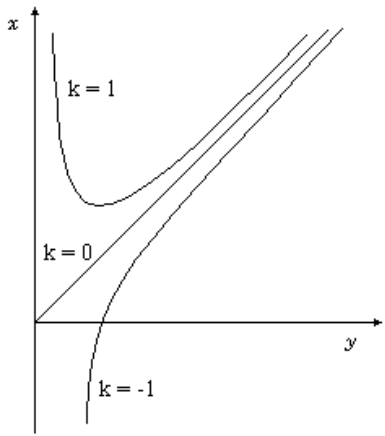, height=6cm}}
\caption{Scale factors relation, that is, the $x~y$ dependence
for the three models.}
\end{figure}
There are several density
values for which the corresponding curves to $k=\pm 1$ models
have vertical asymptotes on the right of vertical axis, limiting
$y$ values below. So, we see that for the KS model, the
scale factor $a(t)$ starts from infinity if $b(t)$ starts from zero.
For the Bianchi I model, the scale factors are always proportional
or even equal. In this situation we don't have an anisotropic model;
in fact, we can easily prove that this model is isotropic by a properly
reparametrization of the coordinates.
For the Bianchi III model, the scale factor $b(t)$
never starts from zero, but has an initial value different from zero
when $a$ is null.
The following plot shows the zones where each model acts (Figure 2).
\begin{figure}[ht]
\centering{\psfig{figure=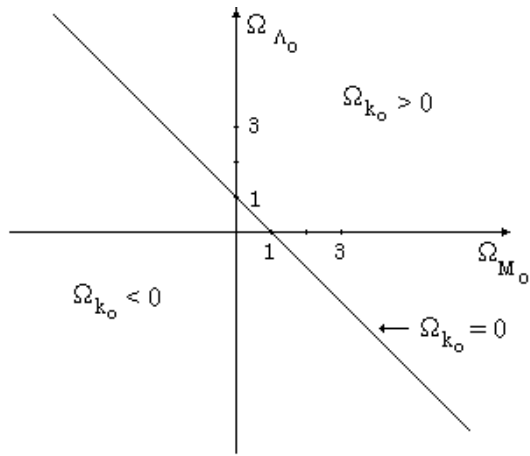, height=6cm}}
\caption{The Kantowski-Sachs model corresponds to the region above
the straight line; the Bianchi III model corresponds the region below
the straight line; the straight line represents
the set of region for the Bianchi I model.}
\end{figure}

Taking into account the analysis given in \cite{Moles}, we may get the
behavior of $y(t)$, since this dimensionless parameter obeys
the same differential equation for these
models and for the standard model.
Now, going back to Figure 1, one can then determine the
$x(t)$ behavior. The plotting below summarizes the several
possibilities for the three models: Kantowski-Sachs, Bianchi I and
Bianchi III models, respectively.
\begin{figure}[ht]
\centering{\psfig{figure=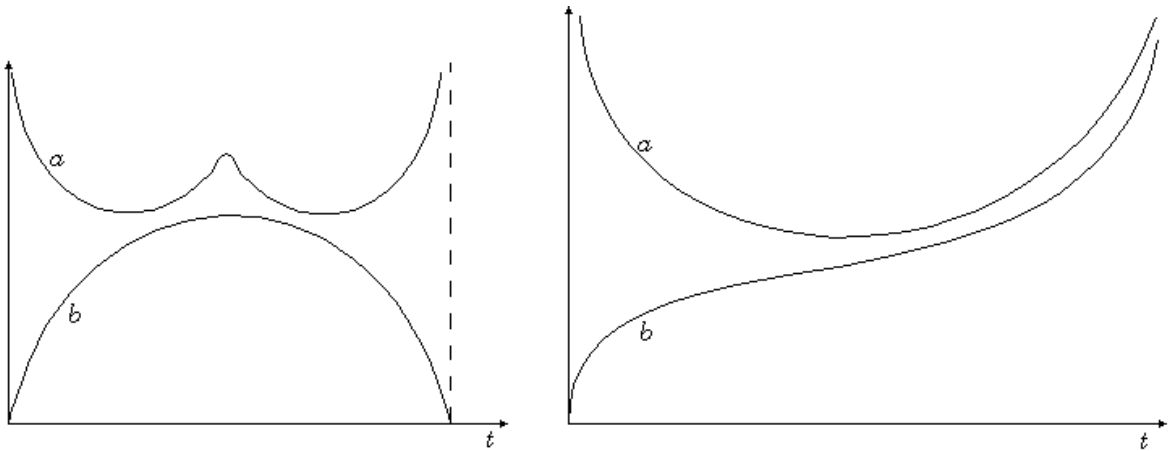, height=6cm}}
\caption{On the left the scale factor for the Kantowski-Sachs
model $(\Omega_{k_{0}}>0)$ when $\Omega_{\Lambda_{0}}<\Omega_{\Lambda_{M}}$.
On the right the scale factor for the Kantowski-Sachs model
$(\Omega_{k_{0}}>0)$ when $\Omega_{\Lambda_{M}}<\Omega_{\Lambda_{0}}
<\Omega_{\Lambda_{c}}$.}
\end{figure}    
\begin{figure}[ht]
\centering{\psfig{figure=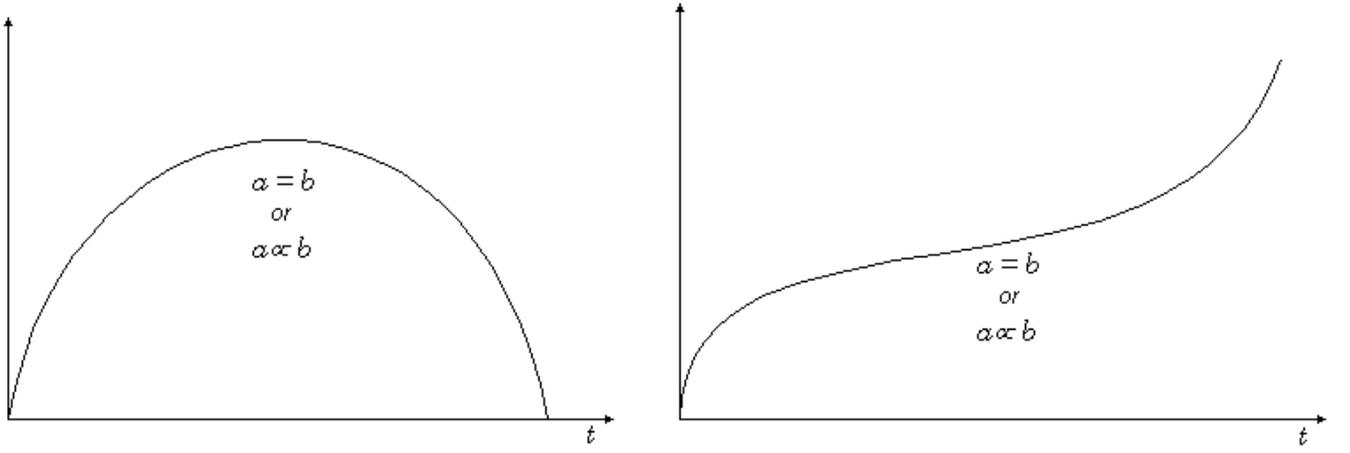, height=6cm}}
\caption{On the left the scale factor for the Bianchi I model
$(\Omega_{k_{0}}=0)$ when $\Omega_{\Lambda_{0}}<0$.
On the right the scale factor for Bianchi I model $(\Omega_{k_{0}}=0)$
when $\Omega_{\Lambda_{0}}\geq 0$.}
\end{figure}
\begin{figure}[ht]
\centering{\psfig{figure=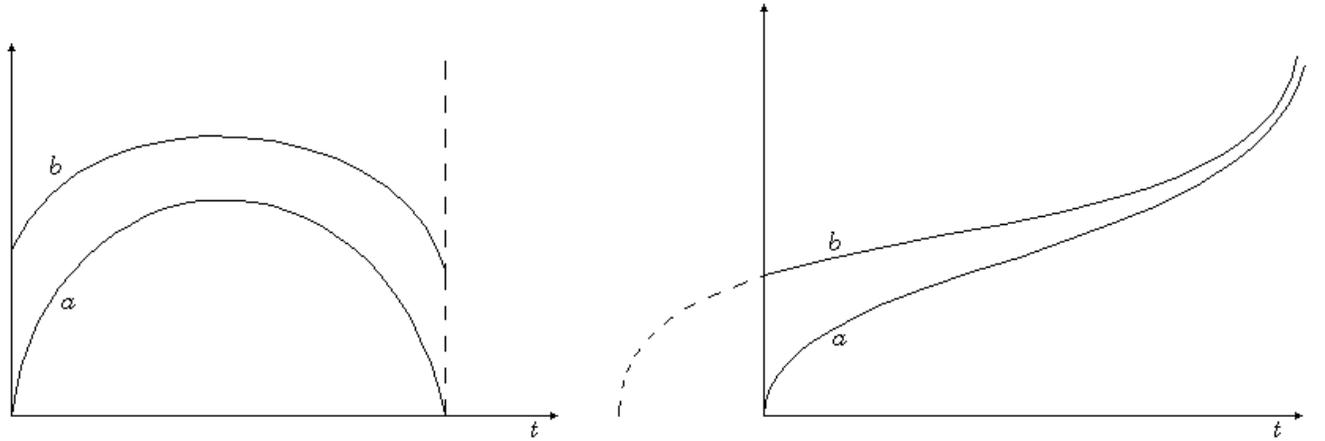, height=6cm}}
\caption{On the left the scale factor for the Bianchi III model
$(\Omega_{k_{0}}<0)$ when $\Omega_{\Lambda_{0}}<\Omega_{\Lambda_{M}}$.
On the right the scale factor for the Bianchi III model $(\Omega_{k_{0}}<0)$
when $\Omega_{\Lambda_{M}}<\Omega_{\Lambda_{0}}<\Omega_{\Lambda_{c}}$.}
\end{figure}

The present technology allows us to ``see" the epoch
of last scattering of radiation at a redshift of about $1000$,
i.e., we can actually observe the most distant information
that the Universe provides. Thus, we observe a great isotropy
from the Cosmic Microwave Background Radiation (CMBR) because
this radiation possesses a near-perfect {\em black body spectrum} 
\cite{Coles}.
The high level of isotropy from this epoch to our days imposes
that the two Hubble factors $H_a$ and $H_b$ must remain approximately
equals from this epoch to the present. In other words, we must impose a high
isotropy level from the last scattering onwards, in our expressions, i.e.,
$${{\Delta H} \over {H_a}}\equiv {{H_a-H_b} \over {H_a}},$$ such that 
$\left| {{{\Delta H} \over {H_a}}} \right| \ll 1$.
We performed several numerical simulations and concluded that the sum
$\Omega_{M_{0}}+\Omega_{\Lambda_{0}}$ must be close to the unity
from above for Kantowski-Sachs and from below for Bianchi III
models\footnote{It is obvious that for Bianchi I model
($\Omega_{M_{0}}+\Omega_{\Lambda_{0}}=1$), with our restrictions, we have 
always $\Delta H/H_a =0$.}. We summarize in the table below the
$\Delta H/H_a$ at $z=1000$, for Kantowski-Sachs and Bianchi III models,
if we supose $H_{a_{0}}=H_{b_{0}}$.

\begin{center}
\begin{tabular}{|c||c|c|c|c|c|}
\hline
\multicolumn{1}{|c||}{} &
\multicolumn{1}{c|}{$\Omega_{M_{0}}$} &
\multicolumn{1}{c|}{$\Omega_{\Lambda_{0}}$} &
\multicolumn{1}{c|}{$\Omega_0$} &
\multicolumn{1}{c|}{$\Omega_{k_{0}}$} &
\multicolumn{1}{c|}{${{\Delta H} \over {H_a}} ~ (\%)$} \\
\hline\hline
K-S & $1$ & $7.46\times 10^{-8}$ & $1-4.974\times10^{-8}$ &
$+7.46\times 10^{-8}$ & $-0.4 \times 10^{-4}$ \\
\cline{2-6}
& $10^{-15}$ & $1$ & $3.33\times10^{-16}$ & $+10^{-15}$ & $-1.0\times 10^{-4}$ \\
\hline
\hline
B III & $1-10^{-8}$ & $9.9\times10^{-9}$ & $1-9.93\times10^{-9}$ & $-10^{-10}$ &
$+9.4\times 10^{-2}$ \\
\cline{2-6}
& $9.99\times10^{-13}$ & $1-10^{-12}$ & $9.997\times10^{-13}$ & $-10^{-15}$ &
$+1.0\times 10^{-4}$ \\
\hline
\end{tabular}
\end{center}

We excluded the situations in that $\Omega_{\Lambda_{0}}$ is close to 
unity, because they are not compatible with the $\Omega_0$ 
values observed today. The two
$\Omega_0$ values $1-6.67\times 10^{-8}$ and
$1-9.93\times 10^{-8}$ for Kantowski-Sachs and Bianchi III models,
respectively, were chosen so they would be both compatible with the estimated
value for our Universe matter density and the restriction
$\left| {{{\Delta H} \over {H_a}}} \right| \ll 1$ on $z=1000$.
These two values for $\Omega_0$ are within the range used in the plots of right
hand side figures (3) and (5), respectively. 

\newpage
{\Large \bf \setlength{\parindent}{0pt} Conclusion}
\vspace{0.5cm}

For the Kantowski-Sachs model, we conclude that if the scale factor $b(t)$
starts from zero, then the scale factor $a(t)$ will start from infinity
and decreases afterwards. When $\Omega_{\Lambda_{0}}<\Omega_{\Lambda_{M}}$,
$b(t)$ reaches the maximum value recollapsing after that. So, $a(t)$ will
reach a relative maximum, when $b(t)$ is maximum, because $x(y)$ has a
relative minimum for $y<1$ (see Figure 1). After that, when $b(t)=0$,
$a(t)$ goes to infinity again.
When $\Omega_{\Lambda_{M}}<\Omega_{\Lambda_{0}}<
\Omega_{\Lambda_{c}}$, the scale factor $b(t)$ grows indefinitely,
giving place to an inflationary scenario. Then, $a(t)$ decreases reaching
a minimum value, and
growing after that indefinitely, and becoming proportional to $b(t)$.
The initial singularity is of a ``cigar" type.

For the Bianchi I model ($\Omega_{k_{0}}=0$), the scale factors $a(t)$
and $b(t)$ are proportional or even equal.
Thus, this model turns out to be an isotropic one.
However, when $\Omega_{\Lambda_{0}}<
\Omega_{\Lambda_{M}}$, $a(t)$ and $b(t)$ reach the maximum and
recollapse after that. And when $\Omega_{\Lambda_{M}}<\Omega_{\Lambda_{0}}<
\Omega_{\Lambda_{c}}$, $a(t)$ and $b(t)$ grow indefinitely after an
inflection.

For the Bianchi III model ($\Omega_{k_{0}}<0$), when $\Omega_{\Lambda_{0}}<
\Omega_{\Lambda_{M}}$, $b(t)$ starts from an initial non vanishing value
$(b(t=0)=b_0>0)$, reaching a maximum and recollapsing after that until
reaches the same value for $t=0$. Also, $a(t)$ has a similar behavior,
but starts from zero and recollapses to zero. When $\Omega_{\Lambda_{M}}<
\Omega_{\Lambda_{0}}<\Omega_{\Lambda_{c}}$, $b(t)$ starts again from
a non vanishing value $(b_0>0)$, growing indefinitely with an inflection.
In this case, $a(t)$ starts from zero and grows indefinitely becoming
proportional to $b(t)$. So, the initial singularity is of a ``pancake" type.

In conclusion, these models undergo isotropisation, except for
the Kantowski-Sachs model $(\Omega_{k_{0}}>0)$ with $\Omega_{\Lambda_{0}}<
\Omega_{\Lambda_{M}}$ and for the Bianchi III $(\Omega_{k_{0}}<0)$ with
$\Omega_{\Lambda_{0}}<\Omega_{\Lambda_{M_{0}}}$.
If we impose that the Universe should be very much isotropic since the 
last scattering epoch ($z\approx 1000$), meaning that the Universe should have
approximately the same Hubble parameter in all directions, we are
led to a matter density parameter very close to the unity at the present time.

\vspace{0.5cm} 
{\Large \bf \setlength{\parindent}{0pt} Acknowledgments}
\vspace{0.5cm}

The authors thank Alfredo B. Henriques, Jos\'e P. Mimoso
and Paulo Moniz for useful discussions and comments.
This work was supported in part
by grants BD 971 e BD/11454/97 PRAXIS XXI, from JNICT 
and by CERN/P/FAE/1164/97 Project.

\end{document}